# Title: Emergence of chiral multi-armed spirals in an open system of migrating cells under continuous cell supply

**Authors:** Masayuki Hayakawa[1,2]†, Biplab Bhattacherjee[1]†, Hidekazu Kuwayama[3], Tatsuo Shibata[1]*

**Affiliations:**

[1] Laboratory for Physical Biology, RIKEN Center for Biosystems Dynamics Research; Kobe 650-0047, Japan.

[2] Department of Mechanical Engineering, Kyoto Institute of Technology; Kyoto 606-8585, Japan.

[3] Faculty of Life and Environmental Sciences, University of Tsukuba; Tennodai, 1-1-1, Tsukuba, Ibaraki 305–8572, Japan.

*Corresponding author. Email: tatsuo.shibata@riken.jp.

†These authors contributed equally to this work

**Abstract:** Chirality organize living and active matter systems into striking collective states, yet the principles that govern chiral ordering in open systems, where elements are continuously added or removed, remain unclear. A mutant strain of *Dictyostelium discoideum* deficient in chemotaxis (KI cells) forms centimeter-scale, clockwise multi-armed spirals. Each arm is a traveling band produced by short-range alignment interactions and guided by a polar-ordered rotating core that encircles the cell source. A subtle clockwise bias in the single-cell migration is amplified by collective ordering into tissue-scale chirality. To uncover the minimal ingredients, we developed an open chiral Vicsek model and identified intrinsic chirality and continuous cell supply as the key factors. Our study establishes a general route by which weak microscopic chirality and sustained material flux generate macroscopic chiral order, offering new insight into chiral patterning in multicellular behaviors.

**Main Text:**

Chirality plays a fundamental role in shaping emergent properties across diverse physical, biological and material systems. Active systems, composed of self-propelled elements that continuously consume energy to sustain motion, exhibit especially rich behaviors when chirality is present (*1, 2*). Chiral active particles, spanning biological molecular motors (*3*) and microswimmers (*4–6*) to synthetic colloidal rotors (*7*), display a variety of intriguing emergent macroscopic behaviors, including chiral edge currents (*5, 7–10*), odd diffusivity(*11, 12*), vortex lattices (*3, 4*), enhanced flocking (*13, 14*), and phase separation into coherently rotating droplets (*13, 14*).

Biological systems are constructed from chiral building blocks such as proteins and DNA. The collective behaviors of these chiral molecular processes (*15, 16*) manifest as active chiral properties in cells (*17–22*), a phenomenon known as cell chirality (*23, 24*). Although cell chirality is considered to underlie left-right asymmetry in organs and bodies (*24–28*), the mechanisms that transmit chiral information from cellular level to the tissue and organ scales remain largely unknown. Insight into this multiscale transfer may come from the self-organization of chiral active matters (*1*).

Most studies of chiral active matter have examined closed systems in which the number of constituent elements is conserved. During morphogenesis, however, cell numbers often change because of proliferation and migration. Neural crest cells, for instance, delaminate from specific regions of the dorsal neural tube and migrate extensively throughout the embryo. In such open systems, which continually exchange matter with their surroundings, tissue-level structures can arise and be sustained by a steady influx of new elements.

**Self-organization of multi-armed spiral structure in mutant *Dictyostelium* cell**

To examine how activity and chirality interact in an open system, we employed a mutant strain of *Dictyostelium discoideum* known as the KI strain. KI cells lack chemotactic ability and therefore fail to aggregate even under starvation (*29*). Instead, they display a distinctive collective behavior: a segregation of cell density that propagates as travelling bands (*30, 31*). Although a chiral bias in KI cells has not been established, wild-type *Dictyostelium* cells exhibit a slight clockwise migration bias (*32*). Therefore, the KI strain offers an ideal model for probing the behavior of active chiral matter in an open system. To investigate this behavior under continuous cell supply, we designed an assay in which cells emerged gradually from defined locations. A pellet of starved KI cells was placed in a 2 mm–radius dimple on the surface of a non-nutrient agar gel (Fig. 1A). Within a few hours the cells began migrating outward, turning the pellet into a persistent cell source. The outward-moving population first formed a dense annulus around the source (Fig. 1B, Movie 1). As time progressed, cells dispersed beyond this annulus, and multiple travelling bands arranged periodically around the circumference produced a centimeter-scale, multi-armed spiral that rotated clockwise, revealing large-scale chiral pattern formation (Fig. 1C, Movie 1).

Velocity analysis revealed that the traveling bands propagated radially outward from the center while rotating clockwise (Fig. 1D). The propagation speed of the bands, $V$, was uniform across positions (Fig. 1E), with the mean speed being 0.363 μm/s.

To assess how cell motion varies across the pattern, we applied particle image velocimetry (PIV) to high-magnification differential-interference-contrast (DIC) images collected in a rectangular window that extended radially from the center to the periphery (Fig. 1F). In the dense annulus immediately adjacent to the dark cell source, velocity vectors pointed predominantly downward, indicating that this region rotates clockwise. Directional alignment was quantified by the polar order parameter $\phi = \langle v/|v| \rangle$, where $v$ is the local PIV velocity.

The high order parameter ($\phi = 0.94$) confirms that this region is polar ordered. Because it rotates clockwise around the field center, we refer to it as the rotating core. The rotating core was already present before the multi-armed spiral appeared (Fig. 1B) and displayed a similar degree of order (fig. S1A,B). In the traveling-band region the order parameter remained high ($\phi = 0.89$), whereas the gaps between bands showed markedly lower alignment ($\phi = 0.6$). These data indicate that KI cells segregate into polar-ordered, high-density traveling bands that advance through a low-density, disordered background (Fig. 1F).

Because the bands emanate from the rotating core and propagate at an almost constant speed across the field, the resulting spiral morphology should be universal, independent of the specific mechanism that generates and transports the bands. A comparable geometry has been analyzed for spiral waves in excitable media (*32, 33*), where the arms follow the involute of a circle. In a polar coordinate $(r, \theta)$ (Fig. 2A) by

$$\theta = \sqrt{(r/a)^2 - 1} - \arccos(a/r) - \Omega t \qquad (1)$$

where $a = V/\Omega$ with $V$ the speed of band propagation and $\Omega$ the angular rotation speed of the pattern (Supplementary Text). Equation (1) can well describe the traveling band shapes traced at individual time points (Fig. 2B, fig. S1C). Over longer timescales, however, the parameter $a$, which characterizes the slope of the curve, decreased (Fig. 2C, blue circle), indicating that the spiral arms became progressively more tilted (Fig. 2D). To identify the cause of this decrease, we measured $V$ and $\Omega$ as functions of time and found that $V$ remained nearly constant whereas $\Omega$ increased gradually (Fig. 2E, fig S2A-C). The slope parameter $a$ obtained from the band shapes matched the ratio $V/\Omega$ (Fig. 2C), confirming that Equation (1) reliably describe our system.

**Alignment interactions promote polar order**

Polar pattern formation is probably driven by direct cell–cell interactions. To characterize these interactions, we examined sparsely seeded KI cells that often formed dynamic clusters undergoing repeated fusion and fission (Fig. 3A, Movie 2). When two solitary cells collided, they tended to align their migration directions (Fig. 3B). In many cases the post-collision separation remained roughly one cell diameter, indicating the formation of two-cell clusters (Fig. 3C). Quantitative analysis showed, the incoming angle $\alpha_{\text{in}}$ is usually larger than the outgoing angle $\alpha_{\text{out}}$ (Fig. 3D), demonstrating a systematic reduction in relative orientation and, therefore, an effective alignment interaction. This short-range alignment interaction likely drives polar pattern formation in KI-cell populations.

**Origin of chirality**

We next investigated the origin of chirality. A previous study showed that wild-type *Dictyostelium* cells tend to migrate clockwise (*34*). We therefore asked whether this single-cell bias underlies the large-scale chirality observed in KI cell populations. To test this idea, we transferred a small number of cells from the rotating core onto a separate non-nutrient agar plate and tracked their trajectories (Movie 3). Because cells readily adhered to one another, we analyzed both solitary cells and small clusters together, following the centroid of each object. Cluster areas ranged from 172 μm² to 1670 μm², with a mean of 1028 μm² (fig. S3A), suggesting that the mean cell number in a cluster was approximately 3.3 assuming the radius of a cell to be 10 μm.

From each trajectory we calculated the angular change in migration direction, $\omega$ (Fig 3E, fig. S3B). The distribution of $\omega$ spanned both positive and negative values, but the mean displacements ⟨ω⟩ was consistently negative in three independent experiments (-3.2304, -2.8102, -1.5956 degrees per min; Fig. 3F). These deviations from zero were statistically significant, demonstrating an intrinsic chirality in the motion of single cell and microscopic cluster. The sign of this bias matches that reported for wild-type cells (*34*).

**Chiral Vicsek model in an open system**

To explain the emergence of the large-scale chiral pattern, we constructed a minimal theoretical model guided by our observations. The microscale polar alignment seen in pairwise collisions, together with the macroscale polar order and traveling bands arising from a disordered background, points naturally to a Vicsek-type description (*35*) augmented with intrinsic cellular chirality (*13, 14*). Because the number of cells outside the source grows over time and eventually exceed $10^6$, we adopt a minimalist approach that captures the essential features by incorporating cell chirality into a Vicsek model in which cells are continuously emitted from a source. In this open chiral Vicsek model, the time evolution of the position $\mathbf{r}_i$ and velocity $\mathbf{v}_i$ of cell $i$ are given by

$$\mathbf{r}_i(t+1) = \mathbf{r}_i(t) + \mathbf{v}_i(t+1) \tag{2}$$

$$\mathbf{v}_i(t+1) = v_0 \mathcal{R}_\alpha \frac{\sum_{j \in R_i} \hat{\mathbf{v}}_j(t) + N_i(t)\eta \hat{\boldsymbol{\xi}}_i(t)}{\left|\sum_{j \in R_i} \hat{\mathbf{v}}_j(t) + N_i(t)\eta \hat{\boldsymbol{\xi}}_i(t)\right|} \tag{3}$$

where $R_i$ is the local neighborhood of the cell $i$ in radius 1 centered at $\mathbf{r}_i$. $N_i$ is the number of cells in $R_i$, $v_0$ is the cell speed, and $\hat{\boldsymbol{\xi}}_i(t)$ is a random unit vector with the strength $\eta$. $\mathcal{R}_\alpha$ is the rotation operation with angle $\alpha$ reflecting the chirality of individual cell with $\alpha = -\omega$ such that ω represents clockwise chirality. To mimic the cell source in the experiment, $2\pi R_0 \kappa$ cells are added and emitted outward from the periphery of the circular source area of radius $R_0$ with the supply rate $\kappa$ per unit length. Throughout this paper, $\eta = 0.3$ and $R_0 = 40$. The source perimeter functions as an effective reflecting boundary. When a cell arrives at this rim from the outside, it is removed, and a new cell is instantly placed at a random location on the rim, oriented to move outward. No other boundaries are imposed in the system.

Our model reproduces the large-scale chiral pattern seen in the experiments. A clockwise-rotating core surrounds the cell source, and multi-armed spirals extend outward (Fig. 4A (i), Movie 4). The core is strongly polar-ordered; the local order parameter is nearly one and the rotation is clockwise (fig. S4A). Outside this core, bands travel clockwise through a low-density, disordered background. The radial density profile peaks inside the rotating core (Fig. 4A (i), bottom). The mean cell orientation $\langle \theta(r) \rangle$, plotted against the distance from the source, reveals a characteristic reorientation sequence. Cells first move directly outward because of the boundary condition, then bend clockwise and begin to circle the source, eventually aligning almost tangentially. After crossing the outer edge of the core and forming spiral arms, the cells gradually turn back toward the radial direction and finally diffuse outward, well away from the source.

At low chirality ω, traveling bands emerge, move radially outward, and form concentric patterns around the cell source without generating a rotating core (Fig. 4A (ii), Movie 5). The radial density profile decays as $\rho(r) \sim 1/r$ independent of $\omega$ (Fig. 4A (ii) bottom). When ω is high, a spot-like pattern appears in which only the clockwise rotating core persists, surrounded by a disordered sea of cells with no spiral arms (Fig. 4A (iii), Movie 6). In this regime the radial density obeys $\rho(r) \sim \exp(-r/\lambda)$, and the characteristic core width follows $\lambda \sim \omega^{-1.5}$, so the spot radius decreases as ω increases (Fig. 4A (iii) bottom). These three different phases, (i) multi-armed spirals, (ii) concentric pattern, and (iii) spots, arise according to the chirality strength $\omega$ and the supply rate $\kappa$ when the noise strength $\eta$ permits optimal alignment (Fig. 4B).

When a multi-armed spiral formed, the traveling bands that radiated from the rotating core could be fitted by the involute of a circle in equation (1) just as in KI cells (fig. S4B). In the simulation however, the slope $a$ increased with time, which contrasts with the experimental result (fig. S4B inset). To clarify this difference, we measured $a$ at a fixed time while varying the supply rate $\kappa$ (Fig. 4C). The slope $a$ grew monotonically with $\kappa$ (Fig. 4C inset, fig. S4C). This trend suggests that, in the experiment, the number of cells in the cell source gradually declined so the effective cell supply rate fell as time passed. Indeed, cell-free patches appeared between the rotating core and the source at later times (fig. S5), consistent with a reduced supply, although the drop in supply rate could not be quantified directly in the experiment.

To test whether the slope $a$ would decrease when the cell-supply rate falls with time, we introduced a time-dependent cell supply defined by $\kappa(t+1) = (1-\gamma)\kappa(t)$, where $\gamma$ is the decay rate. With this modification, the spiral arms still followed the involute of a circle given by equation (1) (Fig. 4D, fig. S4D, Movie 7), yet the slope $a$ now declined over time (Fig. 4E). Furthermore, the normal propagation speed of bands retained a nearly constant speed, while the angular speed $\Omega$ grew steadily, in agreement with the experimental observations (Fig. 4F, fig. S2D). The slope $a$ obtained from the band shapes matched the ratio $V/\Omega$ (Fig. 4E), confirming that equation (1) accurately describes the spiral geometry. This time-dependent supply rate also reproduced the experimental sequence in which the rotating core appears first and the multi-armed spiral develops later (Movie 7).

How does clockwise collective cell migration arise within the rotating core? For unidirectional motion to develop, chirality must couple to some spatial heterogeneity (*36*). As in the case of odd diffusivity (*11*), we considered that this coupling involves the cell chirality $\omega$ and the radial density gradient $\partial_r \rho(r)$ generated by the continuous cell supply. To test this idea, we examined simulation data and measured how the mean orientation of cells at $r$ with respect to the radial direction, $\Theta(r)$, depends on $\omega$ and $\partial_r \rho(r)$ (see Method for the calculation of $\Theta(r)$). When $\omega$ was sufficiently large to sustain a rotation core, the data follows the relation $\tan\Theta \sim \omega\sqrt{|\partial_r \rho(r)/\omega|}$ (fig. S6). This systematic trend supports the idea that the mechanism of the clockwise rotation of the core includes the coupling between cell chirality and the density gradient.

**Discussion**

In our earlier work, we reported nearly periodic traveling bands but detected no clear signs of chirality (*31*). In the present study, detached bands from the rotating core propagate without noticeable rotation. Notably, unlike the excitable media (*32*), traveling bands with free ends never develop into spirals on their own. Together, these observations show that the rotating core is essential for generating the multi-armed spiral pattern formation in this system.

Time-lapse images show that the rotating core expands outward from the periphery of the cell source until it reaches a size that appears to be set by the supply rate. At the surface of this core an instability likely develops, creating undulations that mature into several bands moving clockwise. Because these bands travel almost directly away from the center (Fig. 1D), cells must continuously flow into them along the interface with the core to keep the bands attached, so the bands continually grow and extend outward. When the core was suppressed by lowering the cell density at the source, bands still formed but displayed no clockwise organization (fig. S7). These results indicate that a rotating core fed by a steady cell supply is essential for maintaining the multi-armed spiral pattern of KI cells.

The band-propagation speed $V$ remains constant throughout the field and over time (Fig. 1E, 2E, 4F). Because the bands move tangentially around the circular edge of the rotating core, the pattern's angular speed Ω is set by $V$ and the core radius. The radius gradually shrinks (fig. S8), most likely because the cell-supply rate declines, and this reduction in radius correspondingly raises the angular speed Ω.

The intrinsic chiral bias of cell migration is tiny (about a few degree/min)(*34*), and is usually masked by random motility. Once the cells enter a polar-ordered state, this randomness is averaged out, allowing the bias to emerge as collective chirality. With an average cell speed of about 0.6 μm/sec estimated from the PIV analysis (fig. S9), the expected rotation radius is about 2 mm scale. This matches the size of the rotating core, indicating that a small single-cell chirality can be amplified into centimeter-scale collective chirality.

We demonstrated, using *Dictyostelium*, a rich interplay between chiral active matter and the continuous supply of material, leading to the emergence of large-scale chiral structures such as rotating cores and multi-armed spiral patterns. This collective chirality, observed in a cellular slime mold that retains ancestral traits reflective of early multicellular evolution, it may shed light on the developmental origins of left–right asymmetry in animal body plans.

**Acknowledgments:** We thank Lihao Guo (Shanghai Jiao Tong University) for the assistance during the initial phase of the simulation study and the members of Laboratory for Physical Biology, RIKEN BDR for discussions. We used ChatGPT-4o and 5 to assist with proofreading the manuscript.

**Funding:** This work was supported by Kakenhi grants 22H05170 (TS) and 19K12770 (MH), and core funding at RIKEN Center for Biosystems Dynamics Research.

**Competing interests:** Authors declare that they have no competing interests.

**Data and materials availability:** Source data for all graphs have been deposited and are publicly available at https://doi.org/10.5281/zenodo.17531039.

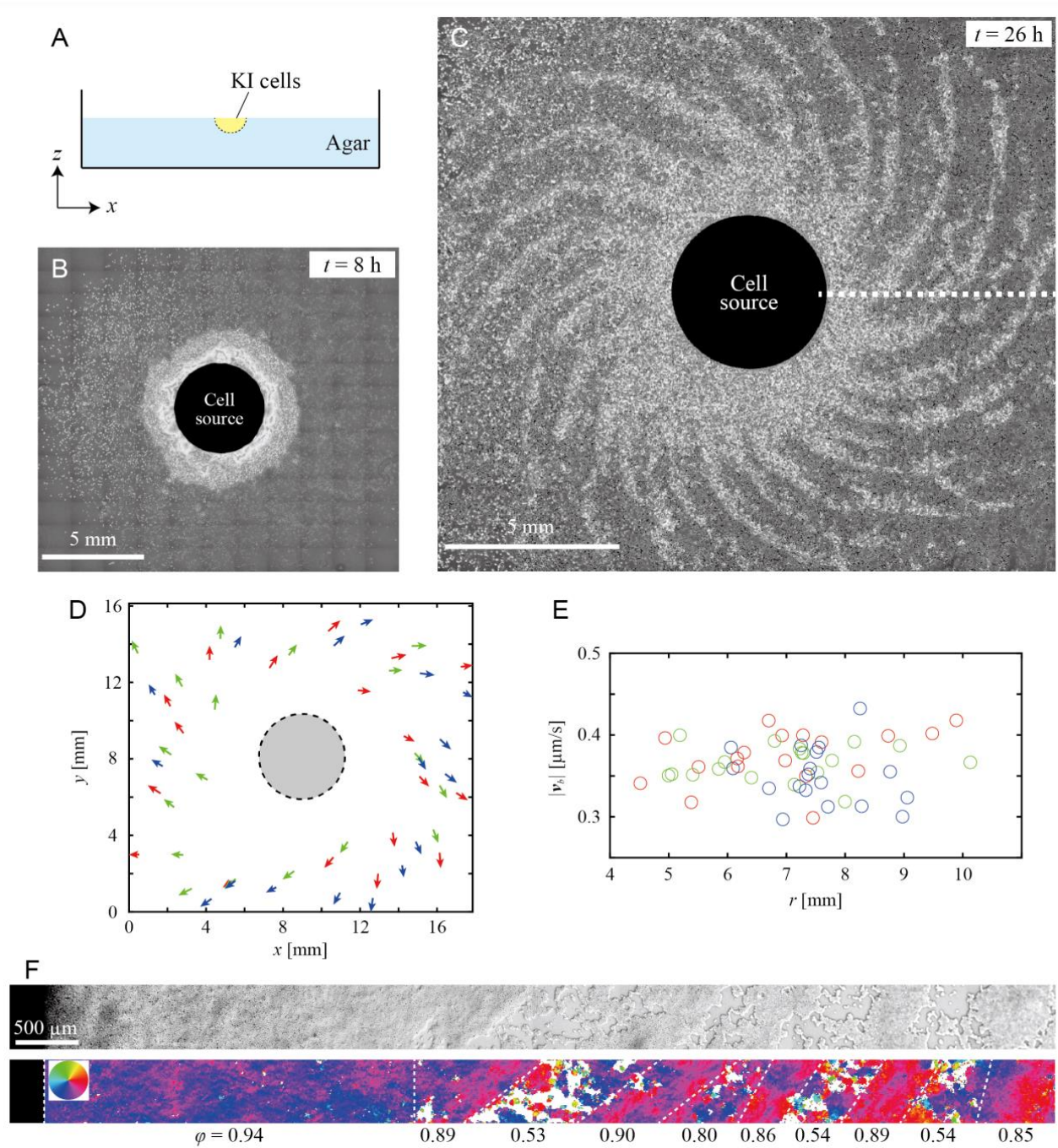

**Fig. 1. Large-scale chiral pattern formation of KI cells.** (**A**) Schematics of experimental setup. (**B**) Formation of an annular structure around the cell source at 8 h. The cell density is high in the brighter region. (**C**) Multi-armed spiral pattern formation with the rotation core around the cell source at 26 h. See also Movie 1. (**D**) Spatial distribution of velocity of traveling bands. The velocity is defined as a propagation normal to themselves per unit time in the three independent experiments shown as different colors. (**E**) Dependence of speed of the traveling bands on the distance ($r$) from the center. (**F**) High-magnification DIC image and particle image velocimetry (PIV) analysis at $t = 26$ h along the narrow window shown as a dotted line in C. The color indicates the direction of motion obtained by PIV. The numbers indicate the polar order parameter $\phi$ obtained in the regions separated by the dotted lines.

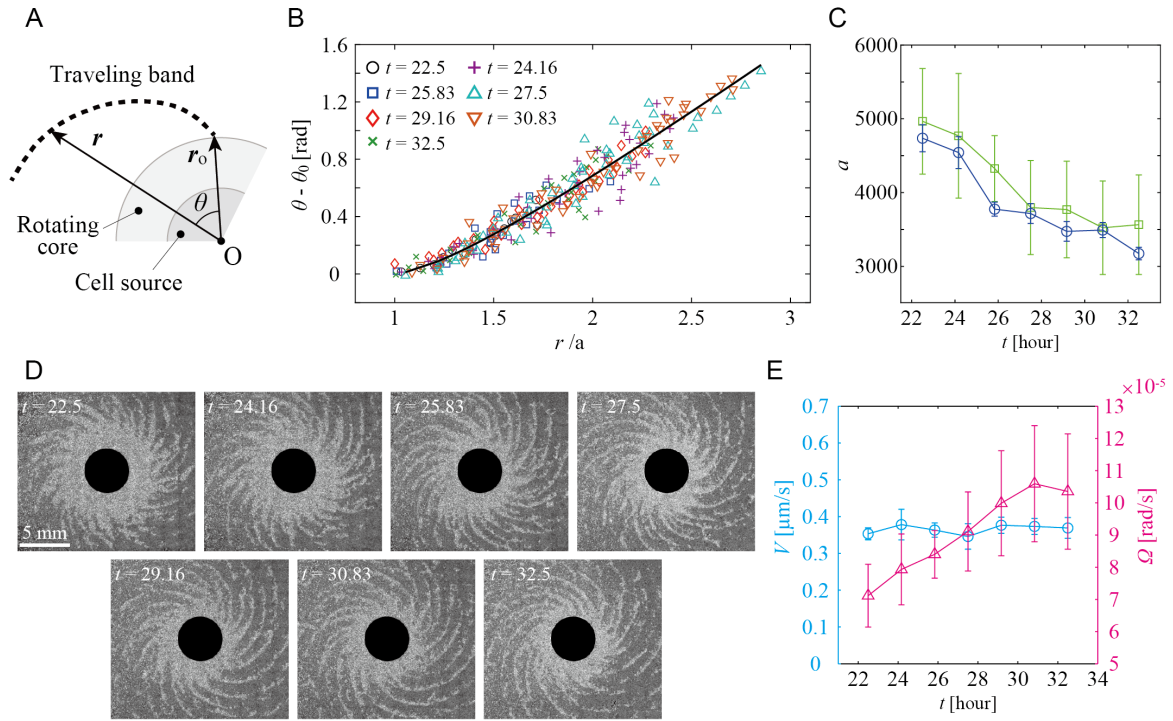

**Fig. 2. Shape of spiral pattern following an involute of a circle.** (**A**) Polar coordinate for the shape measurement of the traveling bands. (**B**) Shape analysis of the traveling band. The shape of traveling band is well described by an involute of a circle given in Eq.(1). (**C**) Time evolution of the slope parameter $a$ obtained by fitting of Eq.(1) to the data (blue) and ratio $a = V/\Omega$ (green). Error bars indicate standard deviations. Sample sizes: $a$ (blue), N = 7, 8, 10, 9, 10, 11, and 12; $V$, N = 8; $\Omega$, N = 17, 20, 18, 17, 17, 15, and 13 (at $t = 22.5 - 32.5$). (**D**) Time-lapse images of spiral pattern showing the patterns become more tilted to the circumferential direction over time. (**E**) Time evolution of the traveling band speed $V$ and the angular speed of pattern rotation $\Omega$, which are the same data used to calculate $a = V/\Omega$ in (C). Error bars indicate standard deviations.

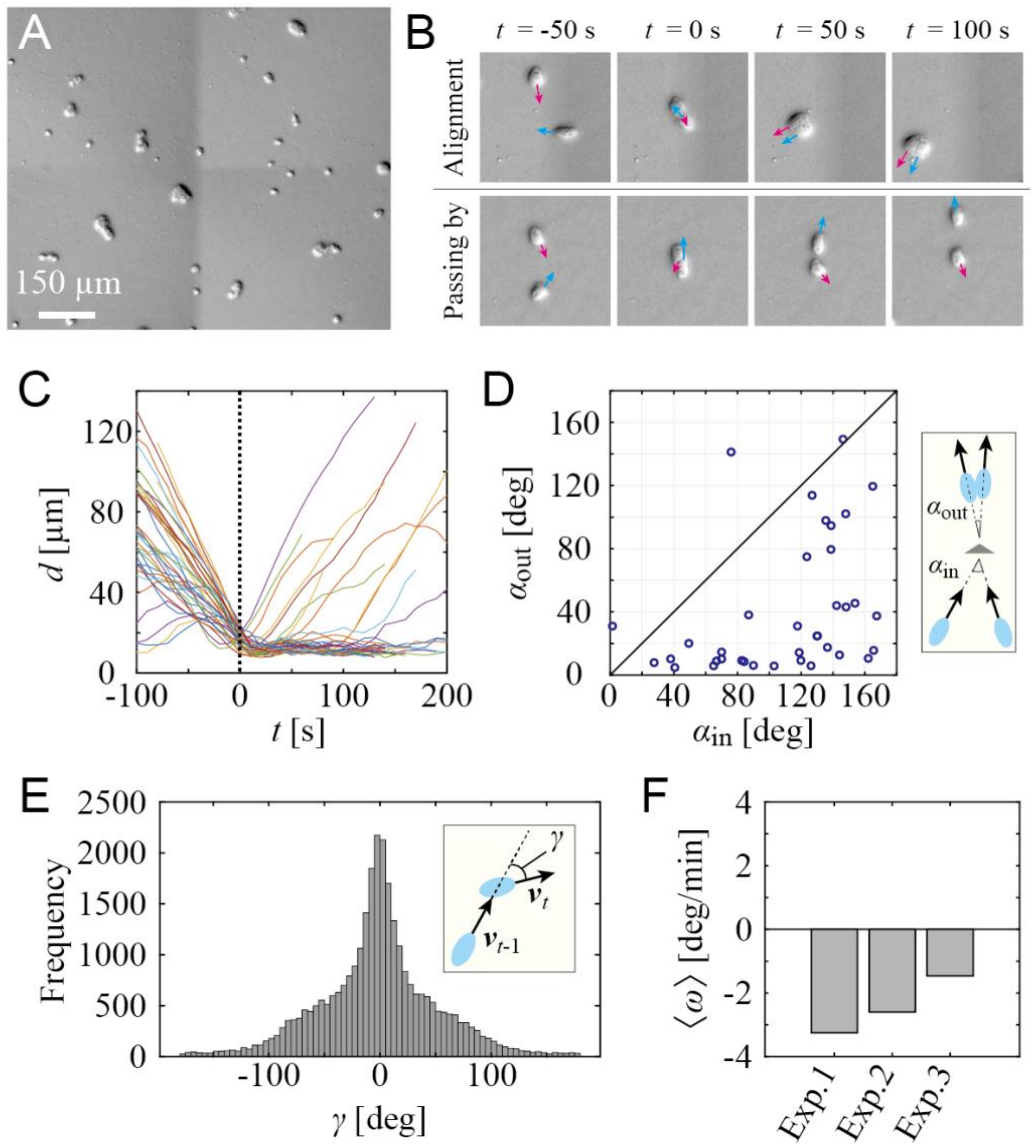

**Fig. 3. Alignment interaction and origin of chirality.** (**A**) Dynamic cluster formation with repeated fusion and fission at low cell density (see also Movie 2). (**B**) Interaction between two cells. Two cells formed a cluster and showed an alignment of migration direction (top) or passed by (bottom) upon collisions. (**C**) Distance between two cells before and after collision. Collision happened at $t = 0$. After the collisions, the distances remained to the cell size in most cases, indicating that they often form cluster. N=37. (**D**) Change in the relative angle of the migration directions upon two cell collision. Most of the point are distributed below the diagonal line ($y = x$), indicating that the outgoing angle tends to be smaller than the incoming angle. N=37. (**E**) Distribution of $\gamma$, the angular change in the migration direction in 30 seconds. Here, we tracked the centroids of single cells and small clusters without distinction. The histogram was constructed from 29819 data points. (**F**) Mean angular velocity $\langle \omega \rangle$ of the KI cells from three independent experiments. $\langle \omega \rangle$ in each experiment was significantly below zero (95% confidence interval entirely negative). Each mean value was calculated from 29819, 17793, and 44340 data points ($\gamma$), respectively.

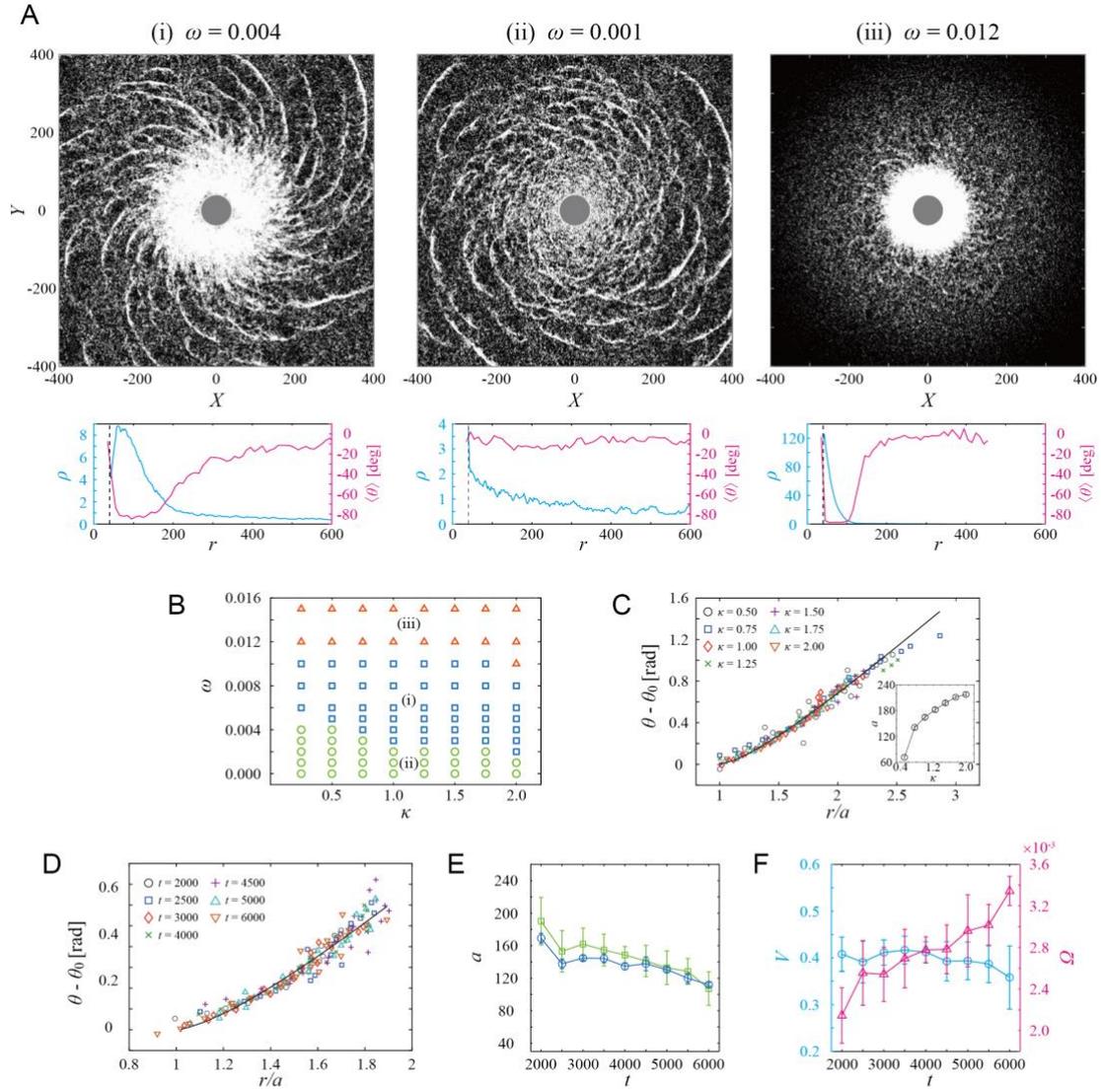

**Fig. 4. Multi-armed spiral pattern formation of open chiral Vicsek model.** (**A**) Three types of pattern dynamics depending on the strength of chirality $\omega$ for $\kappa = 1$. (i) Multi-armed spiral, when chirality is intermediate ($\omega = 0.004$). (ii) Concentric pattern, when chirality is small ($\omega = 0.001$). (iii) Spot, when chirality is large ($\omega = 0.012$). Bottom panels indicate the density profile along the radial direction (blue) and cell orientation relative to the radial direction plotted against the distance from the center. (**B**) Phase diagram for different chirality strength $\omega$ and the supply rate $\kappa$. (**C**) Shape analysis of the traveling band with constant supply rate $\kappa$. The shape of traveling band is well described by an involute of a circle given in Eq.(1). The slope parameter $a$ plotted against the supply rate $\kappa$ (inset). (**D**) Shape analysis of the traveling band with decaying supply rate $\kappa(t)$ with $\kappa(0) = 25$ and $\gamma = 0.006$. $\omega = 0.005$. The shape of traveling band is well described by an involute of a circle given in Eq.(1). (**E**) Time evolution of the slope parameter $a$ obtained by fitting of Eq.(1) to the data (blue) and ratio $a = V/\Omega$ (green). Error bars indicate the standard deviations. Sample sizes: $a$ (blue), N = 11, 9, 15, 11, 13, 12, 11, 10, and 10; $V$, N = 35, 25, 30, 30, 30, 35, 35, 30, and 35; $\Omega$, N = 14, 10, 10, 11, 12, 15, 13, 11, and 14 (at $t = 2000 - 6000$). (**F**) Time evolution of the traveling band speed $V$ and the angular speed of pattern rotation $\Omega$, which are the same data used to calculate $a = V/\Omega$ in (E). Error bars indicate standard deviations.

**Materials and Methods**

Cell culture and preparation of spiral formation

5LP medium (1 liter) was prepared by dissolving 5.2 g of lactose monohydrate (124-00092, FUJIFILM Wako Pure Chemical Corporation) and 5.0 g of Bacto Peptone (211677, Gibco) in 1 L of MilliQ water, followed by autoclaving. For the preparation of the plates, 1.5 g of agar (Powdered Agar for Bacterial Culture Media, Shoei Kanten Co., Ltd) was added to the above before autoclaving.1 mL of *Klebsiella aerogenes* cultured in the liquid 5LP medium for over-night at 37 °C was spread on the 5LP 1.5 % agar plate, and the non-chemotactic *Dictyostelium discoideum*, KI-5 mutant cells were inoculated in the center on the plate. The KI-5 cells were incubated at 21°C for approximately five days. KI-5 (NBRP ID: S00058) can be obtained from National BioResource Project Cellular slime molds (https://nenkin.nbrp.jp).

Following incubation, the KI cells and *Klebsiella* were harvested from the plate using phosphate buffer (PB; 3 mM $Na_2HPO_4 \cdot 12H_2O$, 7 mM $KH_2PO_4$). To remove the *Klebsiella*, the suspension was centrifuged, and the supernatant which contains Klebsiella cells was discarded, followed by the addition of fresh PB. This process was repeated twice to obtain the highly concentrated paste-like KI-5 cell suspension. Approximately 20 µl of the paste-like cell suspension filled in a dimple that was fabricated on the surface of a non-nutrient 1.5 % agar. After a few hours, the cells began to migrate out of the pellet.

The non-nutrient gel having the dimple was prepared on the glass bottom dish (50 mm diameter, D911600, Matsunami Glass Ind., Ltd.). 9 ml of 1.5% agar solution was added and a template of the dimple (hemispherical convex structure on a cover glass) was positioned on surface of the solution. After the solidification, the template was gently removed. The template was fabricated by dropping a UV-curing resin onto a cover glass (24x32 mm, C024321, Matsunami Glass Ind., Ltd.), and allowing it to cure.

Observation of the spiral pattern

All observation was made with the dish upside down to prevent fogging of a lid. All observation was conducted using an inverted microscope (TiE, Nikon, Tokyo Japan) equipped with CMOS camera (DS-Fi3, Nikon, Tokyo Japan). As for the whole spiral observation (Fig. 1B,C and Movie 1), the 11x14 tile images were taken every 10 min using 4x phase-contrast objective. The 1x12 tile images of a narrow window (Fig. 1F) were taken every 15 s using x10 DIC objective.

Image processing of the spiral pattern

Images of the spiral pattern taken by the phase contrast was processed as follows. The process begins by applying a maximum filter (radius is 1 pixel) to the original image. This filter replaces each pixel in the image with the largest value among the neighboring pixels around it. Once the filtered image is obtained, the next step is to subtract it from the original image. Then the difference is added back to the original image. Finally, the contrast tuning and the median filter to reduce the noise was applied. By applying this process, high cell density regions, such as the rotating core and traveling bands, become brighter, improving both visibility and ease of analysis. All process shown here was performed using Image J.

Measurement of the band velocity

To measure the velocity of the traveling band shown in Fig.1D, E, 2D and 4G, first, the normal vectors perpendicular to the local tangents along the traveling band were drawn at a given point $p$ on the band at a specific time point. In the frames before and after that, the points where the normal vectors and the traveling band intersect are defined as $p_{-1}$ and

$p_{+1}$, respectively. Velocity $v$ was obtained by $\frac{p_{+1}-p_{-1}}{2\Delta t}$, where $\Delta t$ is the time interval. For the experimental data, $\Delta t = 15$s, while for the simulation data, $\Delta t = 40$ steps was used. In Fig. 1D, E, the velocities and speeds were obtained at $t = 25$h (sample1), $t = 26.8$h (sample2) and $t = 25$h (sample3).

PIV analysis
    PIV analysis was performed using OpenPIV implemented in MATLAB (The MathWorks, Inc. Natick, MA). A vector field was generated at each time step, and these were then smoothed over time using a 3-point moving average. For a visualization, the color map was created without considering vector intensity, focusing only on orientation. After smoothing, if the vector magnitude was smaller than 0.2 µm/min, the point was regarded as a vacant without cells. The polar order parameter $\phi$ was obtained as $\phi = \frac{1}{N}\left|\sum_{i=1}^{N}\frac{v_i}{|v_i|}\right|$, where $v_i$ is the velocity at lattice $i$ obtained by PIV, and $N$ is the number of lattices in each region.

Characterization of spiral arms as an involute of a circle
    In each snapshot of the spiral obtained in the experiment and simulation, approximately equidistant points were placed manually on the traveling band, and the distance $r$ from center of the cell source O to each point was measured. The vector from O to the contact point between the rotating core and the traveling band was defined as $r_0$. Then the angle $\theta$ of $r_0$ and the vector to each point on the band was measured. Those measurements were performed using Image J. The optimal fitting parameters were determined using a least-squares fit using the analytical form of the involute of a circle, given in Eq. (1).

Quantification of angular speed of pattern rotation
    To obtain the angular speed of the pattern rotation $\Omega$, we first considered the cell density along a circle centered at the cell source with the radius which is larger than the size of rotating core (Fig. S2A). Then, the cell density along the circle is plotted with time as a kymograph (Fig. S2B-D). The high-density regions form lines rising from left to right. The slope of this line gives the angular speed of the pattern. In the simulation, the images for analysis were processed as follows. Particle coordinates were first converted into a two-dimensional histogram. A Gaussian filter was then applied, and a binary matrix was generated by setting grid points with five or more particles to 1. Subsequently, a 3x3 majority filter and a morphological closing operation with a structuring element of radius 3 were applied. Finally, only regions with an area of 2000 or more were retained in the resulting binary image.

Characterization of the alignment interaction
    The cells were carefully scraped from the rotating core with a pipet tip (tip diameter ~ 1 mm) and transferred onto a non-nutrient agar plate. Immediately after the transfer, cells were observed as aggregates. After the 2 hours incubation at 21°C, the aggregate was dissolved due to the migrative activity of individual cells. The behavior of the cells in this state was recorded under the inverted microscope with x10 DIC objective. 2x2 tile image was taken with a time interval of 10 s. A cell-to-cell collision was cropped and the position of both cells were tracked manually. Note that the moment when the cell contours contact each other was defined as $t = 0$. The incoming angle $\alpha_{in}$ was defined as the angle between velocity vectors of each cell at $t = -1$, whereas the outgoing angle $\alpha_{out}$ was defined as the mean angle between the velocities from $t = 5$ to the final observation time. Just after the collision

the cells take time to deform and their velocities fluctuate largely, thus, the definition of $\alpha_{in}$ and $\alpha_{out}$ is not symmetric.

Measurement of the chirality in cell migration

As indicated in previous studies, the chirality, angular velocity of *Dictyostelium* cell migration is not so large, requiring a large number of samples for statistical confirmation. However, due to the tendency of cells to adhere to one another, it is difficult to observe the migration of the single cell with sufficient frequency for robust statistical analysis. Thus, here, the chirality of the migration of cell or small cell clusters were measured. Small amount of the cells was scraped from the rotating core using a pipet tip and transferred onto non-nutrient agar plate. The process that the cells diffuse was recorded using the inverted microscope equipped with the CMOS camera.

The 1x4 tile images were taken every 30 s using 4x phase-contrast objective. This experiment was conducted three times. Obtained images were used for measurement of chirality in cell migration. The trajectories of the cell and cell cluster were automatically tracked by using the TrackMate as a plugin for Image J (National Institutes of Health, USA). The tracker was simple LAP tracker and the detector was Cellpose detector with a pre-trained custom model. To obtain only smaller clusters or single cell trajectories, trajectories of clusters with sizes larger than $1670\ \mu m^2$ were excluded. Additionally, to exclude misrecognitions, trajectories satisfying the following criteria were omitted: total tracking time shorter than 10 minutes, velocity smaller than 0.39 µm/s for all 10 consecutive frames (5 minutes), inclusion of periods with no movement, or an end-to-end distance smaller than 59 µm.

From the trajectories of individual cells obtained in this analysis, we calculated $\gamma$ at each time point and performed statistical analysis on these values. The statistical analysis was based on circular statistics, and the mean and confidence intervals were calculated using the functions circ_mean and circ_confmean in the Circular Statistics Toolbox for MATLAB. We regarded $\gamma$ (i.e., $\omega$) as significantly biased toward negative values when the circular mean was negative and its confidence interval did not include zero.

Analysis of the dependence of the polar direction on the chirality and the density gradient

The direction of polar order $\Theta(r)$ at $r$ the distance from the center is given by $\tan\Theta(r) = \frac{P_\theta(r)}{P_r(r)}$, where $P_\theta(r) = \frac{1}{N(r)v_0}\sum_i v_{i,\theta}(r)$ and $P_r(r) = \frac{1}{N(r)v_0}\sum_i v_{i,r}(r)$. Here, $N(r)$ is the cell number at $r$, and $v_{i,\theta}(r)$ and $v_{i,r}(r)$, are the azimuthal and radial components of the velocity of cell $i$ at $r$.

**Supplementary Text**

Spiral structure described by the involute of a circle

Here, we demonstrate that the morphology of spiral bands can be described by the involute of a circle, as given by equation (1) in the main text (*32,33*). Consider a polar coordinate $(r, \theta)$, where the shape of a spiral band is described by $\theta = f(r) - \Omega t$ with $\Omega$ representing the rotational speed of the spiral pattern around its center. At time $t = 0$, suppose the spiral band intersects the radial line extending from the rotation center at point $A(r, \theta)$. Thus, satisfying $\theta = f(r)$. After a time interval $dt$, the band intersects the same radial line at point B, satisfying $\theta = f(r + dr) - \Omega dt$. Consequently, the radial displacement AB is expressed as $dr = \Omega\, dt / f'(r)$. At $t = dt$, the spiral band intersects the circle of radius $r$ at point C. The arc length AC along the circle is given by $r\Omega dt$. Thus, the angle $\phi$ formed between the spiral band and the circle of radius $r$ satisfies

$$\tan \phi = 1/rf'(r) \tag{1}$$

Next, the distance between two successive spiral bands at $t = 0$ and $t = dt$ is given by $Vdt$ with $V$ the normal propagation speed of the spiral band. Hence, we obtain

$$\sin \phi = V/r\Omega \tag{2}$$

Combining Eqs. (1) and (2), we derive the following differential equation

$$f'^2 = (\Omega/V)^2 - 1/r^2 \tag{3}$$

Solving this differential equation yields the explicit form of $\theta = f(r) - \Omega t$, as presented by equation (1) in the main text.

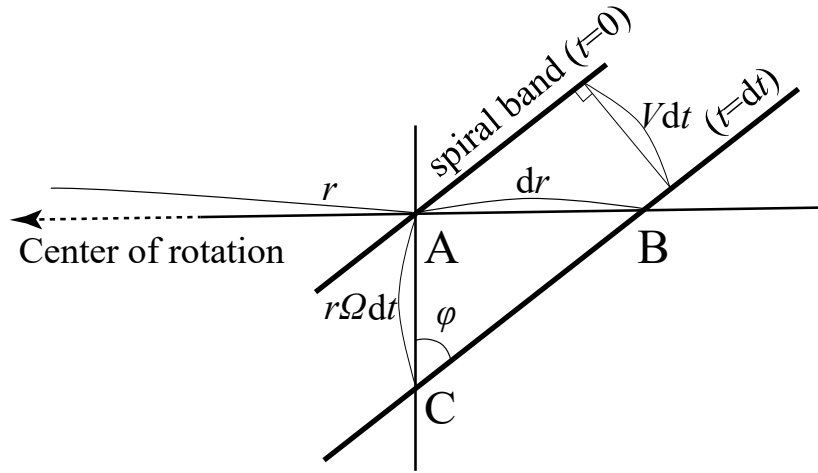

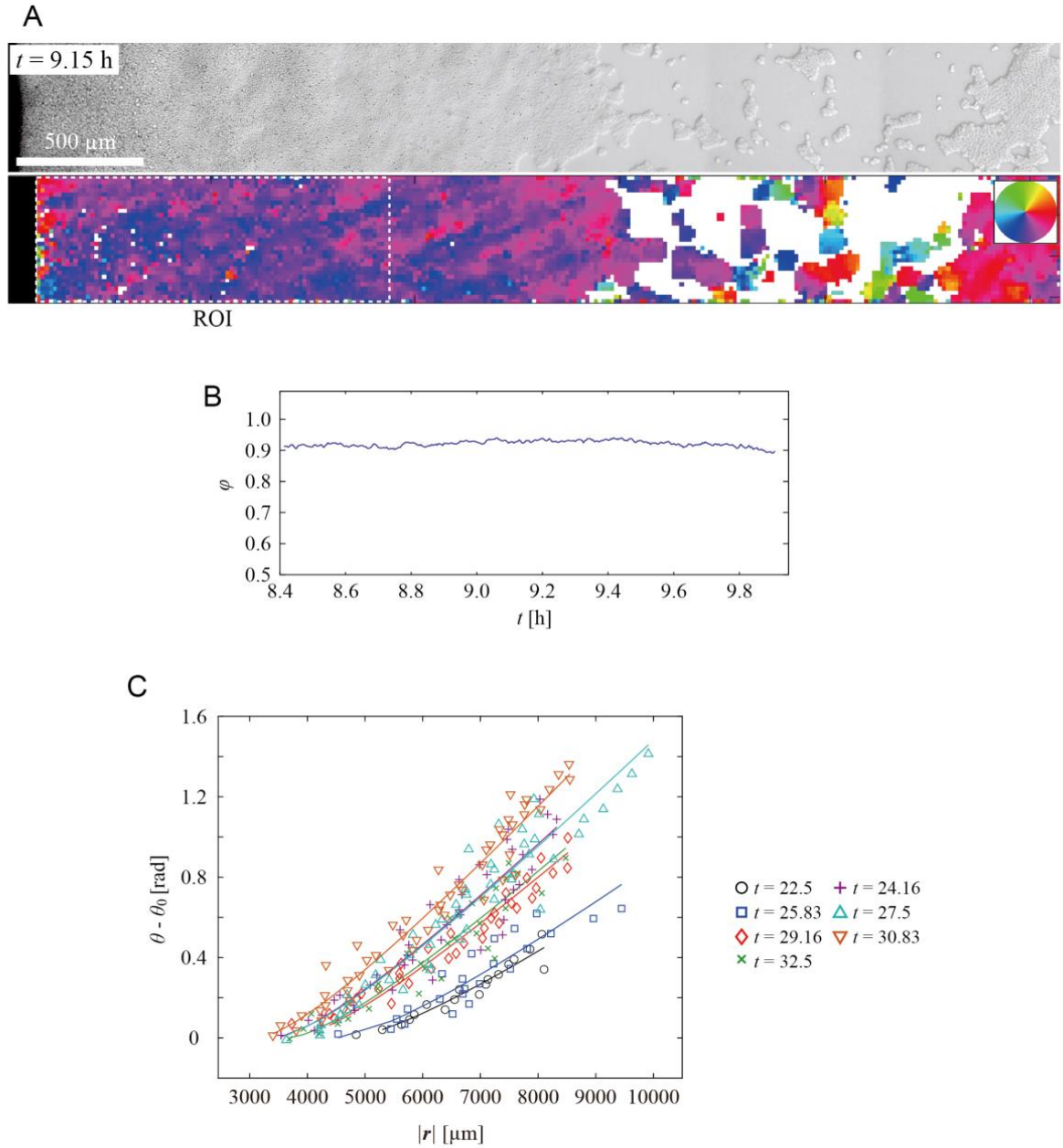

**Fig. S1. Formation of rotating core with polar order prior to the emergence of multi-armed spiral and shape of spiral arms**. (**A**) A high-magnification DIC image and the particle image velocimetry (PIV) analysis along the narrow window as in the case of Fig.1C at 9.15h. The color indicates the direction of motion obtained by PIV. (**B**) Time series of the polar order parameter $\phi$ obtained in the region indicated in A. The high value of around $\phi \approx 0.9$ indicates that the rotating core is in a polar-ordered state. (**C**) Temporal evolution of the spiral shape shown in Fig.2B.

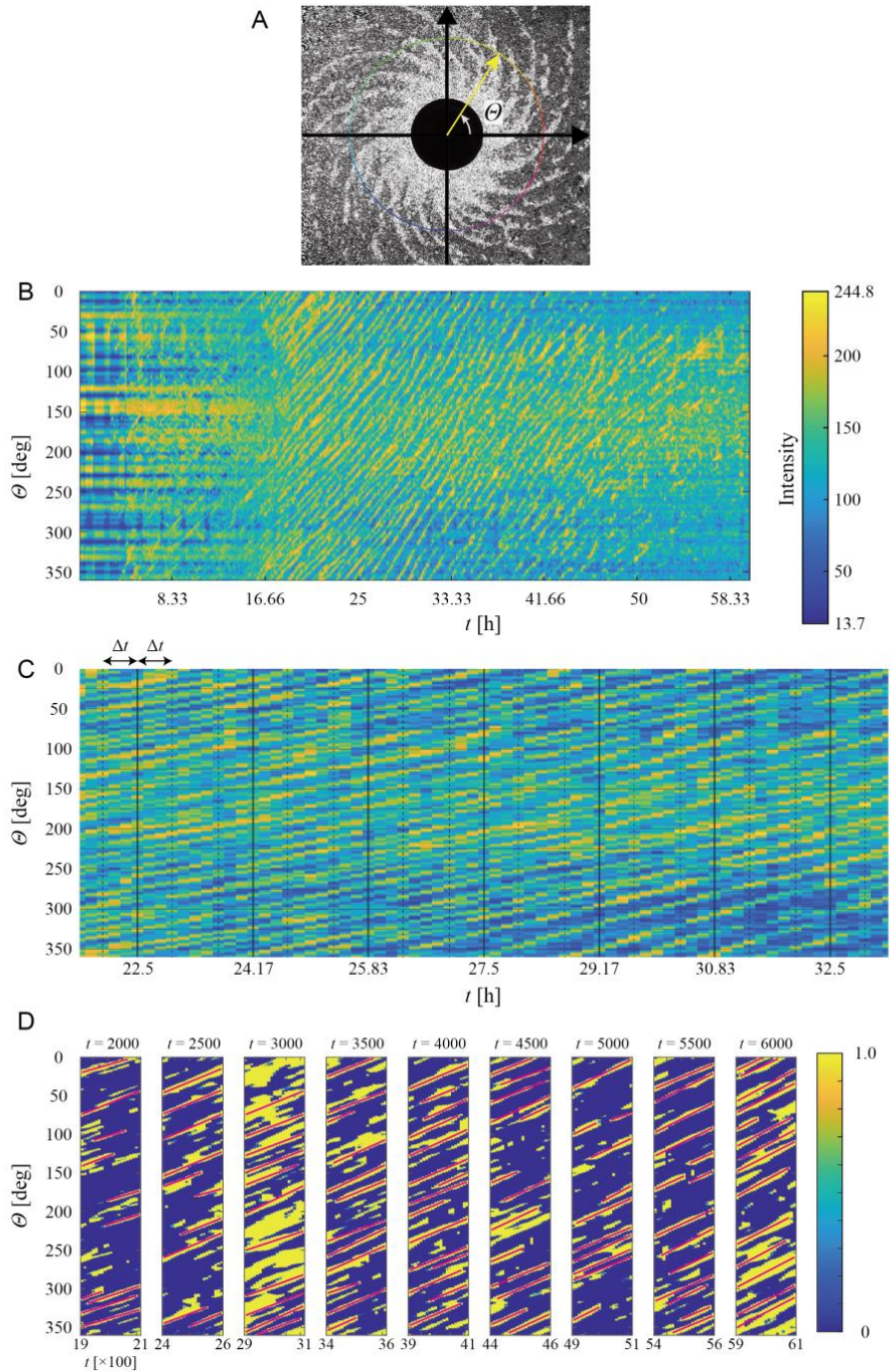

**Fig. S2. Quantification of angular speed of pattern rotation.** (**A**, **B**, **C**) Experiment and (**D**) simulation of open chiral Vicsek model. (**A**) Cell density profile considered along the circle. (**B**) The cell density in the angular direction $\Theta$ along the circle in A plotted over time. (**C**) The magnified image in B. The slop of the high-density region gives the angular speed of pattern rotation $\Omega$. (**D**) Cell density profile along the circle at each time point. The slope was measured within a range of $\pm 100$ around each time point (magenta line). See Materials and Methods for the detailed description of the identification of $\Omega$.

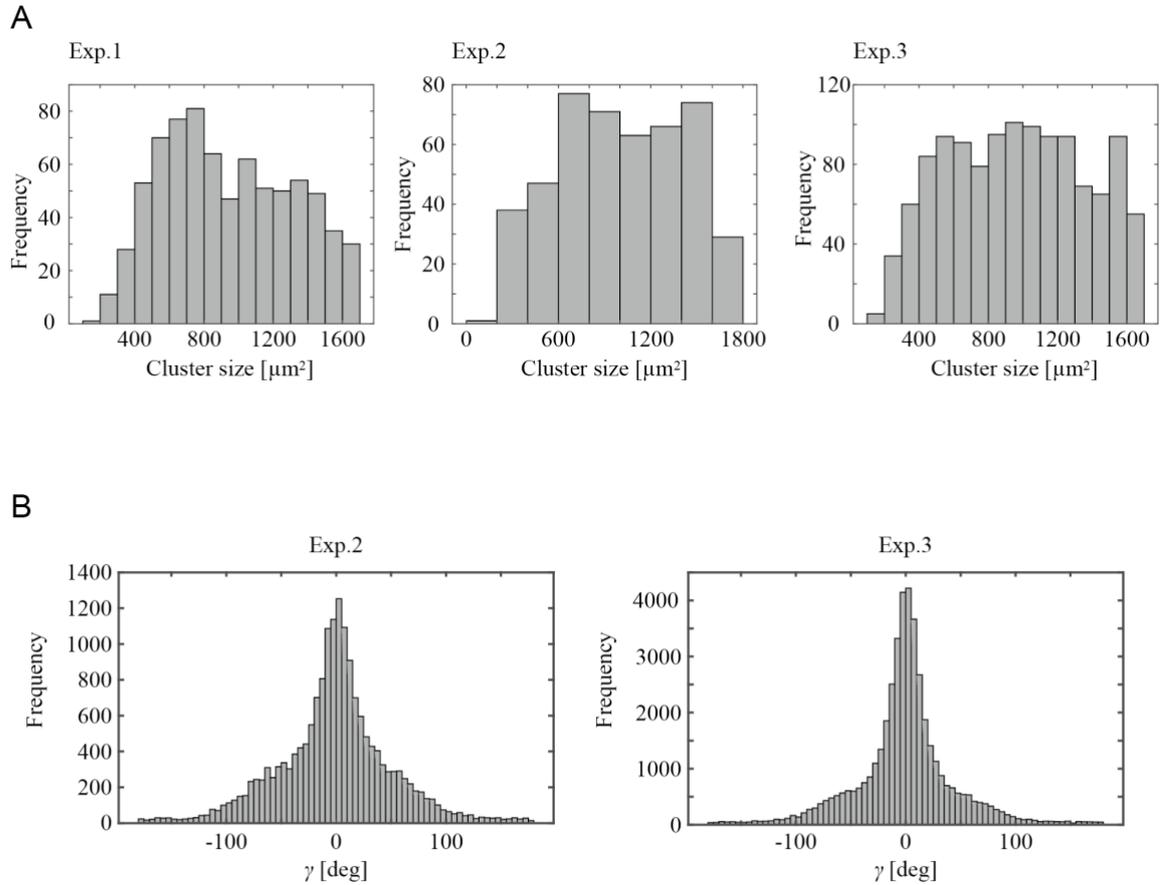

**Fig. S3. Origin of chirality** (**A**) Distribution of cell cluster size in the cell chirality assay. Distribution of cell cluster size in the three independent experiments. The size distribution of the tracked clusters ranged from 172 μm² to 1670 μm², with a mean of 1028 μm² for experiment 1, suggesting that the mean cell number in a cluster was approximately 3.27 assuming the radius of a cell to be 10 μm. (**B**) Distribution of angular change $\omega$ in the cell chirality assay. Distribution of angular change in the migration direction in 30 seconds in the 2nd and 3rd experiments in addition to the 1st experiment shown in Fig.3E.

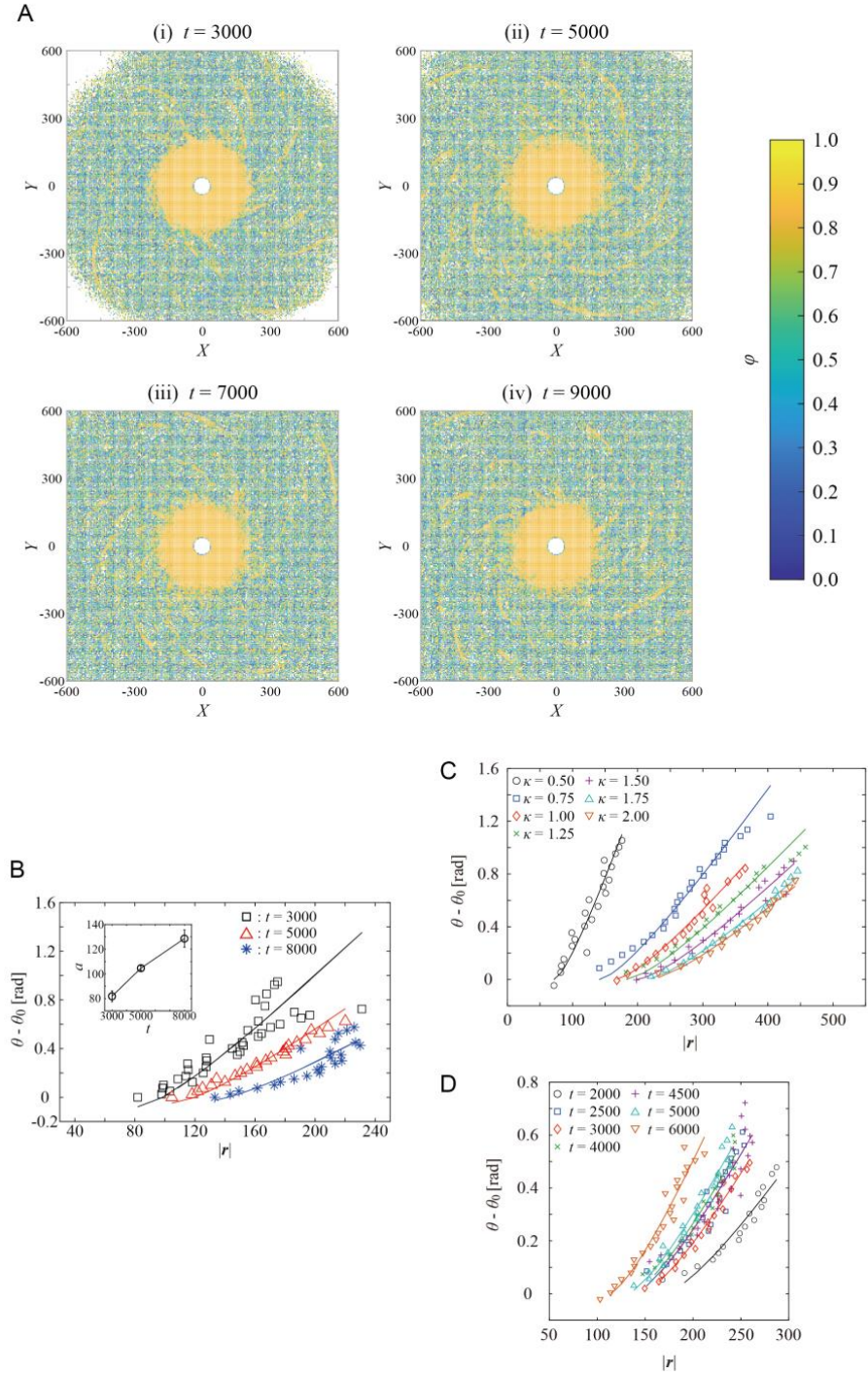

**Fig. S4. Spatiotemporal dynamics of open chiral Vicsek model.** (**A**) Spatial distribution of local polar order parameter in open chiral Vicsek model. (**B**) Temporal dynamics of the spiral shape parameterize by slope $a$ for the case when the cell supply rate was constant. (**C**) Dependence of spiral shape on the supply rate $\kappa$ shown in Fig 4C. (**D**) Temporal evolution of the spiral shape in the case of the time-dependent supply rate $\kappa$ as shown in Fig.4D.

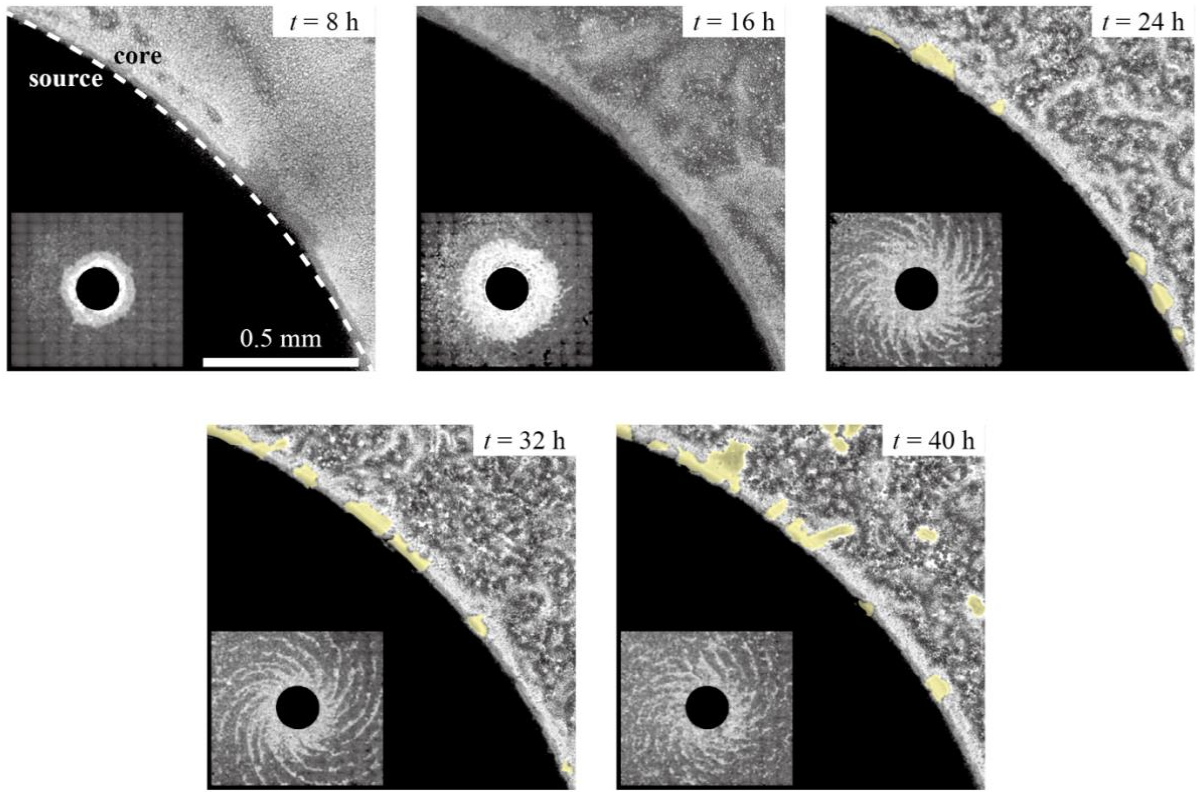

**Fig. S5. Appearance of cell-free regions near the core–source boundary.** Over time, cell-free vacant spaces appear due to a decrease in the cell supply from the source. The yellow areas indicate the vacant spaces. The image in the bottom right of each panel shows the overall pattern at the corresponding time point.

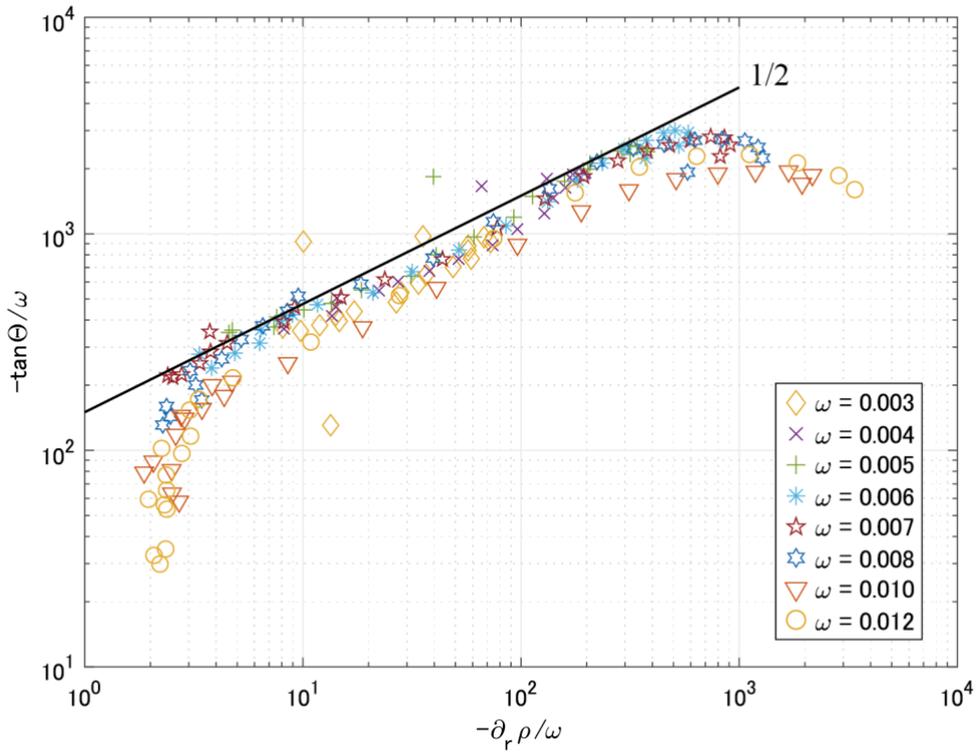

**Fig. S6. Dependence of polar order direction on the density gradient.** The polar order direction $-\tan(\Theta)/\omega = -P_\theta/(P_r\omega)$ plotted as a function of the density gradient in the radial direction $-\partial_r\rho(r)/\omega$, showing a power-law dependence with exponent $1/2$, suggesting that $\tan\Theta \sim \omega\sqrt{|\partial_r\rho(r)/\omega|}$. Here, $P_\theta = \langle v_{i,\theta}\rangle/v_0$ and $P_r = \langle v_{i,r}\rangle/v_0$ where $v_{i,\theta}$ and $v_{i,r}$ are the tangential and radial components of velocity $\mathbf{v}_i$ of cell $i$.

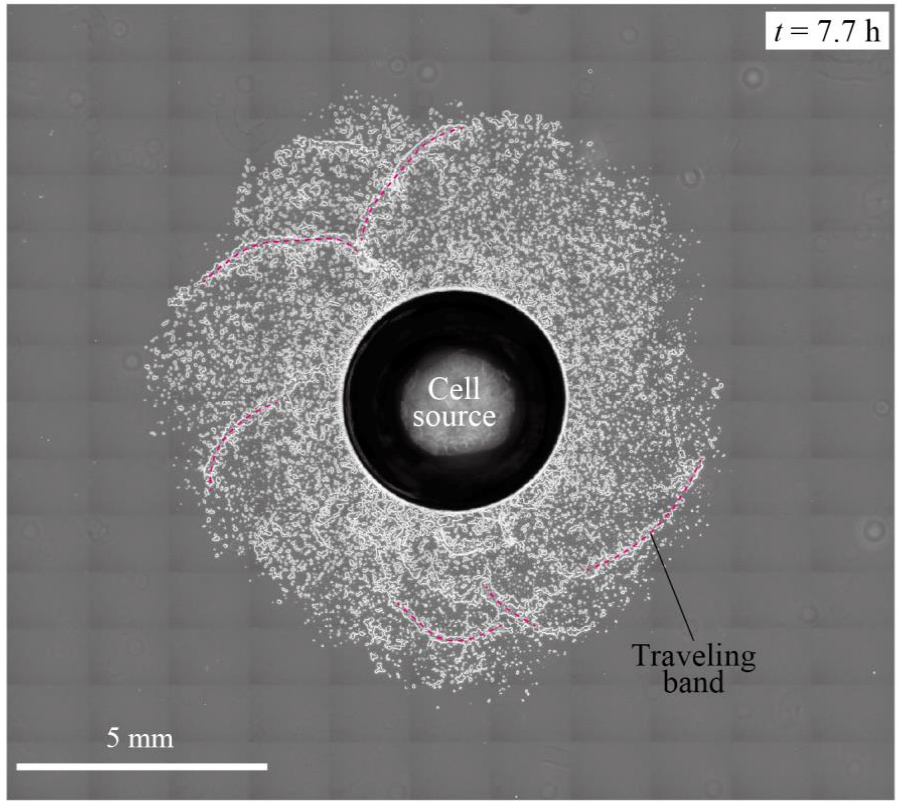

**Fig. S7. Traveling bands without rotating core.** When 3 µl of the dense cell suspension was used, the traveling bands were formed without forming rotating core. The bands were propagating in the random direction without forming a multi-armed spiral pattern.

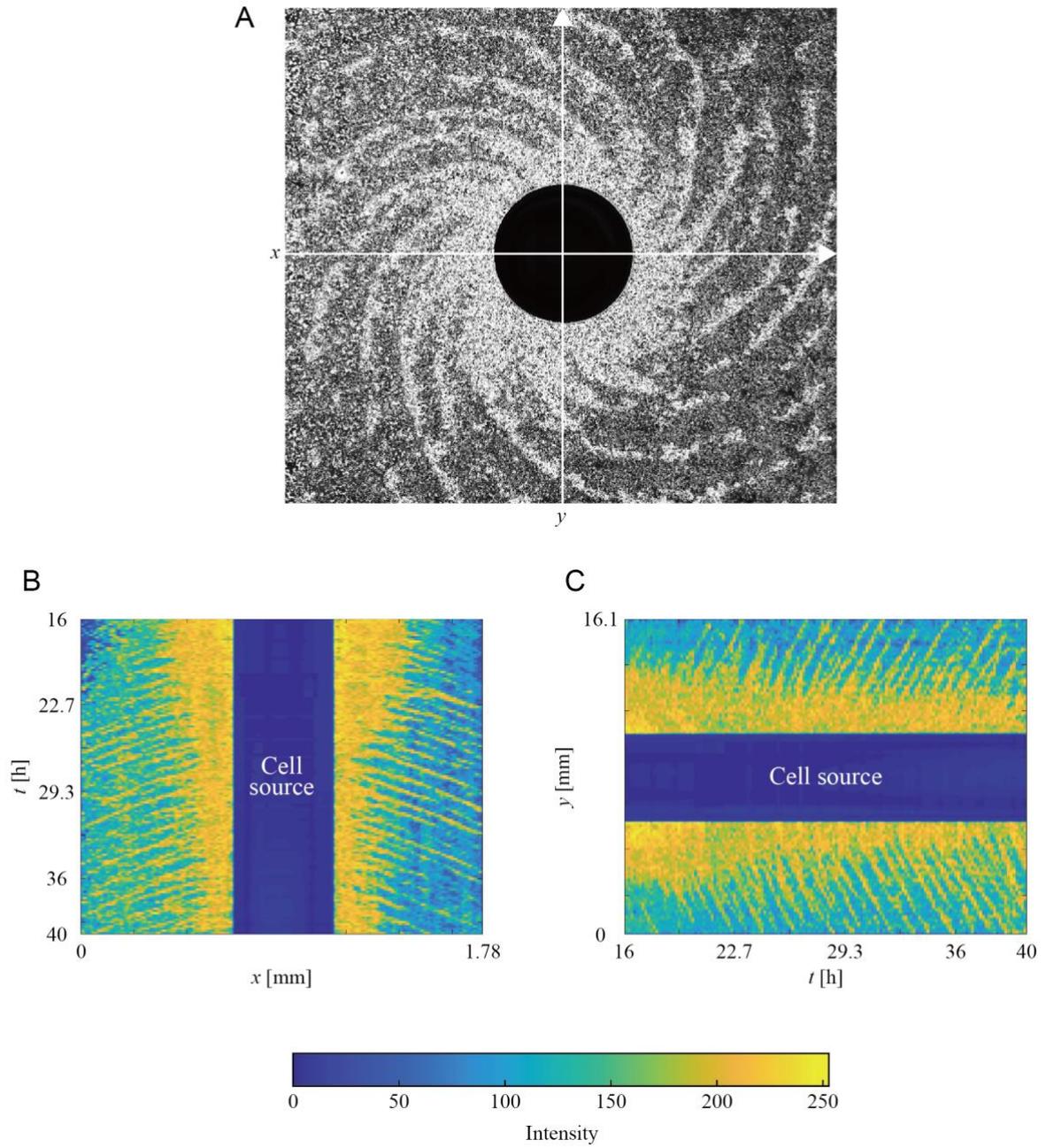

**Fig. S8. Shrink of the size of the rotating core.** (**A**) Cell density profile consider along the $x$ and $y$ axies. (BC) Cell density profile along the $x$ axis (**B**) and $y$ axis (**C**). The width of the high cell density area around the cell source showed a decay over time.

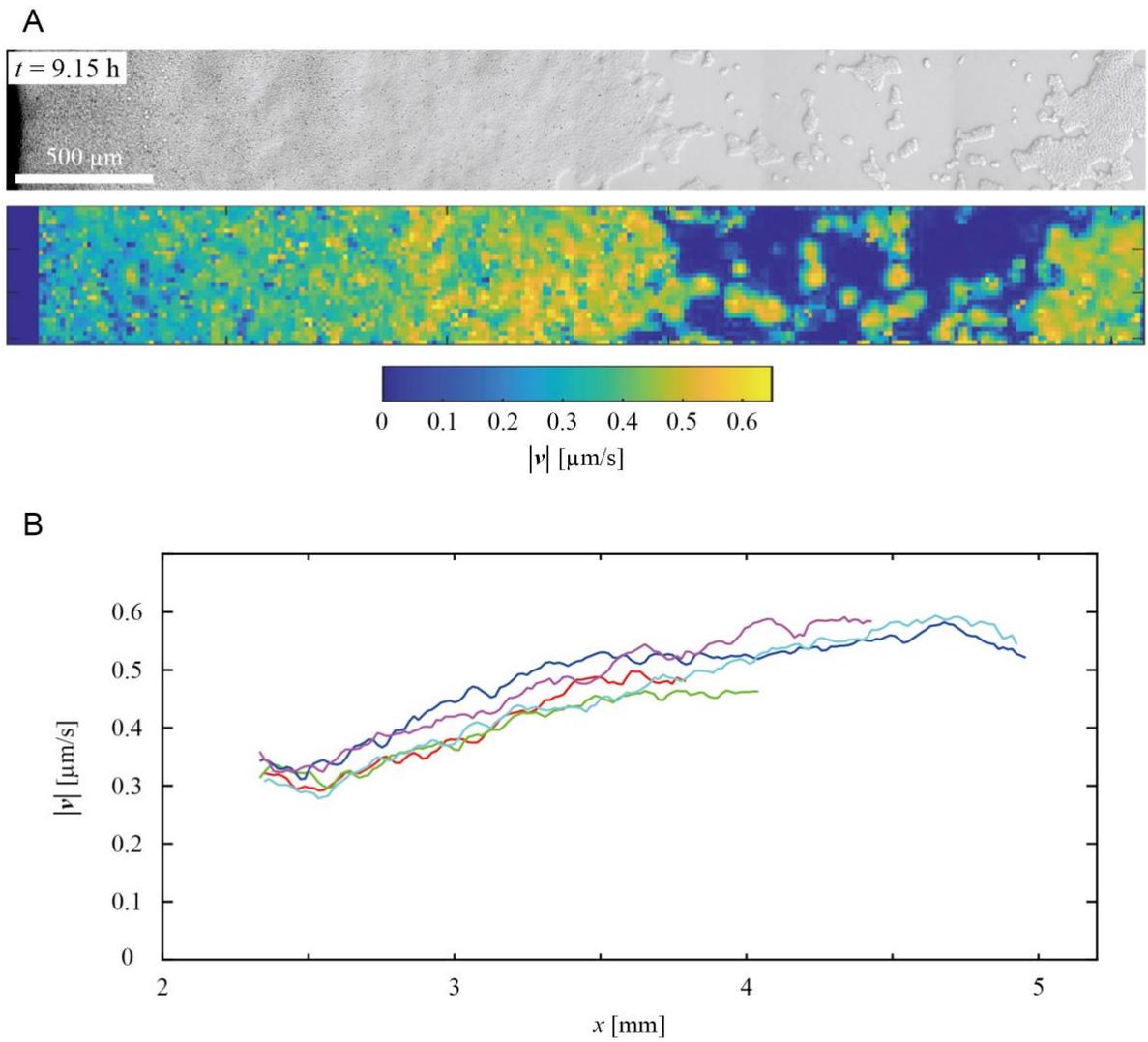

**Fig. S9. Cell speed estimated from PIV analysis.** (**A**) PIV analysis performed at 9.15h in a narrow window as in the case of Fig. 1F and fig. S1A. (**B**) Estimation of cell speed from the PIV analysis. The estimated values are distributed from 0.3 to 0.6 µm/sec.

**Movie S1.** Large scale chiral pattern formation with multi-armed spiral of KI cell.

**Movie S2.** Cell-cell interaction assay at low cell density. Cells were transferred from the rotating core onto a non-nutrient agar plate. Cells frequently formed dynamic clusters undergoing repeated fusion and fission.

**Movie S3.** Chirality assay of cells and cell clusters measured at low cell density. Cells were transferred from the rotating core onto a non-nutrient agar plate.

**Movie S4.** Large scale chiral pattern formation with multi-armed spiral and rotating core in open chiral Vicsek model. $\omega = 0.004$.

**Movie S5.** Concentric pattern formation in open chiral Vicsek model. $\omega = 0.001$.

**Movie S6.** Spot formation in open chiral Vicsek model. $\omega = 0.012$.

**Movie S7.** Large scale chiral pattern formation with multi-armed spiral and rotating core in open chiral Vicsek model with decaying supply rate $\kappa(t)$ with $\kappa(0) = 25$ and $\gamma = 0.006$. $\omega = 0.004$.